\documentclass[amsmath,amssymb,aps,superscriptaddress]{revtex4-2}

\usepackage{graphicx}% Include figure files
\usepackage{bm}% bold math
\usepackage{hyperref}% add hypertext capabilities
\hypersetup{
    colorlinks=true,
    linkcolor=blue,
    filecolor=magenta,      
    urlcolor=blue,
    citecolor=blue,
}
\usepackage{xcolor}% To get more colors
\usepackage{geometry}

\newtheorem{example}{\textbf{Example}}

\begin{document}

\title{Comment on ``Synchronization dynamics in non-normal networks: the trade-off for optimality''}

\newcommand{\nuphys}{\affiliation{Department of Physics and Astronomy, Northwestern University, Evanston, IL 60208, USA}}
\newcommand{\nico}{\affiliation{Northwestern Institute on Complex Systems, Northwestern University, Evanston, IL 60208, USA}}

\author{Takashi Nishikawa}
\email{t-nishikawa@northwestern.edu}
\email{tnishi2012@gmail.com}
\nuphys
\nico

\author{Adilson E. Motter}
\nuphys
\nico

\author{Louis M. Pecora}
\affiliation{Code 6392, U.S. Naval Research Laboratory, Washington, DC 20375, USA}

% \date{\today}

\begin{abstract}
A recent paper by R. Muolo, T. Carletti, J. P. Gleeson, and M. Asllani [\href{https://doi.org/10.3390/e23010036}{Entropy \textbf{23}, 36 (2021)}] presents a mainly numerical study on the role of non-normality in the synchronization of coupled periodic oscillators, deriving apparent contradictions with the existing literature.
Here, we show that their conclusions are artifactual due to a misinterpretation of the master stability function (MSF) formalism and confirm that the existing literature is correct.
We also point to a broader existing literature in which research considered in this paper was correctly addressed numerically, analytically, and experimentally.
\end{abstract}

\maketitle

Ref.~\cite{muolo2021synchronization} uses a network of diffusively coupled oscillators to question the validity of existing approaches to study synchronization in non-normal systems.
Unfortunately, the results presented in the paper are either incorrect or known in the literature (despite the lack of proper attribution), as we show below.

\subsection*{Misrepresentation of the MSF in favor of an inadequate approach}

\begin{quote}
``The MSF relies on the computation of the (real part of the maximum) Lyapunov exponent and, thus, in the case of time-dependent systems, it does not possess the full predictability power that it has in the autonomous case (fixed point in/stability).'' \cite{muolo2021synchronization}
\end{quote}
This argument forms the basis of the paper and, unfortunately, is compromised.
First, the Lyapunov exponents are real by definition, and thus there is no reason to take the real part of them, as repeatedly claimed in the paper. Second, the Lyapunov exponents can be computed for time-dependent systems, even for chaotic ones \cite{eckmann1985ergodic}. Third, the MSF applies to time-dependent states just as well as it applies to time-independent ones \cite{pecora1998master,pecora2000synchronization}. 
Motivated by these three misconceptions, the paper proceeds with:
\begin{quote}
``For this reason, we will use a homogenization method, whose validity is limited to a specific region of the model parameters, allowing for us to transform the linearized periodic case problem into a time-independent one [26] [Ref.~\cite{sanders2007averaging} here].''
\end{quote}
The proposed homogenization is an unnecessary complication and unjustified approximation for a problem that already admits an exact formulation and clean solution. The homogenization is used to create what the paper calls an ``autonomous version of the MSF,'' which is specious since the MSF (being a Lyapunov exponent \cite{pecora1998master,fink2000three}) is a time-independent quantity by definition.
Crucially, the use of this method as proposed in the paper invalidates the conclusions of any subsequent stability analysis {\it even within the linear regime}. 
Specifically, for networks of coupled periodic oscillators, the homogenization method replaces $A(t)$ in the time-periodic linearized system $\dot{x} = A(t)x$ by its time-average $\bar{A} := \frac{1}{T} \int_0^T A(t)dt$ over the period $T$ of the function $A(t)$.
As we show in the following example, stability is generally not preserved under this operation in the presence of non-normality (which \textit{is} the scenario considered in Ref.~\cite{muolo2021synchronization}).
\begin{example}
Consider the analytically solvable two-dimensional system from Ref.~\cite{wu1974note} defined by
\begin{widetext}
\begin{equation}
A(t) = \begin{bmatrix}
-1 - 9\cos^2(6t) + 12\sin(6t)\cos(6t), & 12\cos^2(6t) + 9\sin(6t)\cos(6t) \\
-12\sin^2(6t) + 9\sin(6t)\cos(6t), & -1 - 9\sin^2(6t) - 12\sin(6t)\cos(6t)
\end{bmatrix},
\end{equation}
\end{widetext}
which is periodic with period $T = \pi/6$.
The trajectory of the system starting from the initial condition $x(0) = x_0$ can be written explicitly as $x(t) = \Phi(t,0) x_0$, where $\Phi(t,0)$ is the state transition matrix of the system, given by
\begin{align}
\Phi(t,0) = \frac{1}{5} \begin{bmatrix}
e^{2t}\phi(t) + 2e^{-13t}\psi(t), & 2e^{2t}\phi(t) - e^{-13t}\psi(t) \\[1mm]
e^{2t}\psi(t) - 2e^{-13t}\phi(t), & 2e^{2t}\psi(t) + e^{-13t}\phi(t)
\end{bmatrix},
\end{align}
and $\phi(t)$ and $\psi(t)$ are $\pi/3$-periodic functions defined by
\begin{align}
\phi(t) &:= \cos(6t) + 2\sin(6t) \quad\text{and}\quad \psi(t) := 2\cos(6t) - \sin(6t).
\end{align}
Due to the exponential factor $e^{2t}$ in $\Phi(t,0)$, all solutions will diverge exponentially as $t \to \infty$, and thus the equilibrium state $x = 0$ is unstable in this system.
On the other hand, the time-average of $A(t)$ produced by the homogenization method in Ref.~\cite{muolo2021synchronization} takes the form
\begin{align}
\bar{A} = \begin{bmatrix}
-\frac{11}{2} & 6 \\
-6  & -\frac{11}{2} 
\end{bmatrix}.
\end{align}
The eigenvalues of this matrix are $-11/2 \pm 6i$, and thus the equilibrium $x=0$ is stable for the corresponding homogenized system $\dot{x} = \bar{A} x$. 
\end{example}
It follows that this homogenization method cannot be used to determine the stability of time-dependent systems (which should not be confused with justified uses of averaging over fast oscillations in the study of systems with well-separated time scales \cite{sanders2007averaging}).
Naturally, there are valid approaches to eliminate the explicit time dependence, such as the use of  Poincar\'e sections \cite{chicone2006ordinary}. Those approaches would lead to the same (correct) conclusions obtained by directly considering the MSF for the original system.

After incorrectly concluding that existing approaches are not adequate, the paper proposes ``using the numerical technique that is based on a spectral perturbation concept that is known as the pseudo-spectrum.'' 
By doing so, it confounds the identification of optimal networks with their robustness to structural changes.
Indeed, the pseudo-spectrum \cite{trefethen1991pseudospectra,trefethen2005spectra} is defined for a given $\varepsilon>0$ as the set of all possible eigenvalues that can be realized for the Laplacian matrix of the network when \emph{the network structure is modified} by a perturbation whose norm is  less than $\varepsilon$.
Thus, the pseudo-spectrum is adequate for studying the extent to which the stability properties of a network withstand changes in the network structure but not for studying the stability of a state against dynamical perturbations.
Importantly, this type of robustness for optimal networks had already been studied rather systematically in Refs.~\cite{nishikawa2017sensitive,fish2017construction}. Moreover, inspired by the theoretical analysis in Ref.~\cite{nishikawa2010network}, the robustness of synchronization in empirical optimal networks had been studied experimentally and numerically already in Ref.~\cite{ravoori2011robustness}.

\subsection*{Misinterpretation of the properties of optimal networks}

In the analysis of optimal networks, Ref.~\cite{muolo2021synchronization} states:
\begin{quote}
``For a family of models (e.g., R\"{o}ssler, Lorenz, etc.), whose stable part of the MSF has a continuous interval where the (real part of the) Laplacian eigenvalues can lie, it has been proven that they maximize their stability once the coupling network satisfies particular structural properties. Such optimal networks should be directed, spanning trees and without loops [5,6] [Refs.~\cite{nishikawa2006maximum,nishikawa2006synchronization} here]. These networks have the peculiarity of possessing a degenerate spectrum of the Laplacian matrix and laying in the stability domain that is provided by the MSF. The Laplacian degeneracy is also often associated with a real spectrum or with considerably low imaginary parts when compared to the real ones [11,12] [Refs.~\cite{marrec2017analysing,trefethen2005spectra} here].''
\end{quote}
Several corrections are in order. For the class of systems considered in Refs.~\cite{nishikawa2006maximum,nishikawa2006synchronization}, the optimal networks have been proved to have identical Laplacian eigenvalues (except for the one that is necessarily zero). As a corollary, it has also been proved that they must be real (since eigenvalues of real matrices come in conjugate pairs).
However, having real eigenvalues is by no means evidence that a network is optimal or close to being optimal: the Laplacian eigenvalues of undirected networks are \textit{always} real, and yet a sufficiently large undirected ring network of oscillators with a finite MSF stability region is unsynchronizable \cite{pecora1998master}.
Moreover, with the exception of globally coupled unweighted networks, no undirected network is optimal \cite{nishikawa2006maximum,nishikawa2010network}. 
In addition, while the optimal networks include a class of ``directed, spanning trees'' (which are loopless by definition), the cited references (Refs.~\cite{nishikawa2006maximum,nishikawa2006synchronization}) do not suggest that they are required to be of that form. On the contrary, it has long been shown that the set of all ``[s]uch optimal networks'' is quite large and comprises a diverse range of complex network structures (including structures with loops) \cite{nishikawa2006maximum,nishikawa2010network,ravoori2011robustness,fish2017construction}.

These equivocated interpretations, combined with the homogenization-based numerics (already shown to be problematic) performed on {\it two} networks \cite{muolo2021synchronization}, were followed by the conclusion that
\begin{quote}
``This [homogenization] method allows for proving that networks whose Laplacian matrix exhibits a spectrum that lacks an imaginary part are the most optimal.''
\end{quote}
The part of this statement that is correct
(that lacking the imaginary part is necessary for the optimality)
was already established rigorously in Refs.~\cite{nishikawa2006maximum,nishikawa2006synchronization} under general conditions (i.e., not only for specific network realizations). Moreover, Ref.~\cite{nishikawa2010network} introduced a measure of synchronizability that accounts for both the real and the imaginary parts of the Laplacian eigenvalues.

\subsection*{Failure to apply the existing solution to the problem}

The focal point of Ref.~\cite{muolo2021synchronization} is optimal networks and transient growth induced by non-normality in such networks.
Unfortunately, there are three additional crucial errors in this portion of their analysis. 
First, the MSF is applied as formulated in Ref.~\cite{pecora1998master} assuming that the Laplacian matrix is diagonalizable (i.e., has a complete basis of eigenvectors), which is an assumption that was justified in the context of Ref.~\cite{pecora1998master} but violated in the context of Ref.~\cite{muolo2021synchronization}. 
Indeed, as conveyed already in the title of Ref.~\cite{nishikawa2006synchronization}, this is not the case for \textit{optimal} networks (with the only exception of directed unweighted star networks and superpositions of multiple such networks of the same size; see Theorem 6 in Ref.~\cite{nishikawa2006maximum}).
A proper formalism to address such cases, based on the Jordan decomposition, was presented in Refs.~\cite{nishikawa2006maximum,nishikawa2006synchronization}.
While the eigenvalue condition for linear stability turns out to be the same as in the original formulation of the MSF, this formalism also yields a coupled set of equations that governs the transient growth behavior that Ref.~\cite{muolo2021synchronization} claims to have discovered.
Using those equations, the transients in directions transverse to the synchronization manifold can be calculated analytically (not just numerically) for networks of coupled phase oscillators.
Specifically, for each Jordan block of size $s>1$ associated with an eigenvalue $\lambda$ for the Laplacian matrix, the $k$th perturbation mode $\eta_k(t)$ of the transients in the corresponding generalized eigenspace can be expressed as 
\begin{equation}
\eta_k(t) = \sum_{m=0}^{k-1} \frac{(-1)^m \eta_{k-m}(0) \sigma^m}{m!} \cdot t^m e^{-\sigma\lambda t}
\end{equation}
for $k = 1, 2, \ldots, s$, where $\sigma$ is the global coupling strength constant \cite{nishikawa2006maximum,nishikawa2006synchronization}.
In the case of directed chain networks, the Laplacian matrix has a Jordan block of maximum possible size, $s=n-1$, where $n$ denotes the network size.
We note that directed chain networks include as a special case the only optimal network structure explicitly considered in Ref.~\cite{muolo2021synchronization} (the network in Fig.~1b for $\varepsilon=0$).
We further note that the disregard for the coupled eigenmodes that result from the proper MSF formulation also affects their analysis of non-optimal networks that are non-diagonalizable, including degenerate networks considered in Fig.~A1.

Another crucial problem with Ref.~\cite{muolo2021synchronization}'s analysis of the networks that are ``most optimal'' is that different networks are compared for a fixed choice of the coupling strength parameters $D_{\phi}$ and $D_{\psi}$ (diffusion coefficients in their system). 
Thus, a network could synchronize for a larger range of parameters and yet be misdiagnosed as less synchronizable depending on the coupling strength used.
This is not how the MSF-based synchronizability analysis was formulated or intended to be used \cite{pecora1998master,barahona2002synchronization}.
Also, the comparison between networks in Ref.~\cite{muolo2021synchronization} does not constrain the (weighted) in-degrees, effectively causing the more non-normal cases (smaller $\varepsilon$ in Figs.~1a and 1b) to correspond to `sparser' networks. The conclusion that in the specific case they consider the synchronization basin of attraction is ``reduced for the non-normal network as compared to the normal one'' might be true based on previous studies, but it is also plausible that the basin is reduced in sparser networks. 

Therefore, aside from the technical problems above, the conclusions that could be drawn from Ref.~\cite{muolo2021synchronization} are perforce less general than the existing ones (they would even depend on the coupling strength). This is at variance with the actual conclusions presented in the paper:
\begin{quote}
``Based on the well-known Master Stability Function, it has been shown that directed tree-like networks are optimal for models with a discontinuous interval of the Laplacian spectrum in the stability range of MSF. Here, we have extended such results, proving that they are generally independent of the dynamic model.''
\end{quote}
Indeed, Ref.~\cite{muolo2021synchronization} does not show the optimality of any network not yet known in the literature, nor extend existing optimality results, nor prove the independence of any result on the dynamic model used.
Similarly unjustified is the assertion in Ref.~\cite{muolo2021synchronization} that they ``have extended the idea of non-normal dynamics to the case of non-autonomous synchronization dynamics,'' given that the notion of non-normality used is simply the conventional definition of matrix non-normality applied to the Laplacian matrix of the network, disregarding the role played by the (non-autonomous) Jacobian matrix.
Thus, the paper does not provide a non-autonomous extension of the non-normality concept.

\subsection*{Confusion between infinitesimal and finite-size perturbations}

\begin{quote}
``Our results clearly show that networks previously thought to be optimal regarding synchronization are not such, but, on the contrary, the stability of the associated synchronous solution is quite fragile to small perturbations, which makes their role in the synchronization dynamics apparently different from what was previously intuited in the literature [5,6] [Refs.~\cite{nishikawa2006maximum,nishikawa2006synchronization} here]. Additionally, the non-normality makes standard techniques, such as the Master Stability Function, fail by a large amount.'' \cite{muolo2021synchronization}
\end{quote}

These conclusions are incorrect not only because of the methodological errors described above but also because they are 
ultimately based on the numerical simulations presented in Figs.~4 and 5 of Ref.~\cite{muolo2021synchronization}.
It is from these numerics that they draw an apparent contradiction with previous results rigorously established using the extension of the MSF formalism in Refs.~\cite{nishikawa2006maximum,nishikawa2006synchronization}.
In reality, there is simply no contradiction: their numerics are based on \emph{finite-size} perturbations, which are inadequate to draw definite conclusions about the stability of the synchronization trajectory against \emph{infinitesimal} perturbations.
It is an elementary fact that, if finite-size perturbations are allowed, then sufficiently large perturbations can always be used to cross the boundary of the synchronization basin of attraction whenever this basin is not global (as commonly observed in oscillator networks).

To fully appreciate the issue, we first recall that the MSF is the maximum Lyapunov exponent of the synchronization trajectory associated with perturbations transverse to the synchronization manifold \cite{pecora1998master}.
A strictly negative MSF thus indicates that, after an infinitesimal perturbation transverse to the synchronization manifold, the perturbed trajectory converges back to the synchronization manifold with an asymptotic average exponential rate given by the MSF.
We note that this is generally not equivalent to the transverse asymptotic stability of the corresponding attractor in the synchronization manifold, which is a stronger notion of stability defined as follows.
An attractor in the synchronization manifold is said to be transversely asymptotically stable if, for any given $\varepsilon>0$, there exists a $\delta>0$ such that after any transverse perturbation of magnitude $<\delta$, the trajectory stays within distance $\varepsilon$ of the attractor for all $t$ and converges to the attractor as $t\to\infty$.
The difference between the two notions of stability can be observed in a phenomenon known in the literature as \emph{bubbling} \cite{ashwin1994bubbling,heagy1995desynchronization,venkataramani1996transitions}, which can occur in the synchronization of chaotic oscillators.
However, for the case of periodic attractors (as considered in Ref.~\cite{muolo2021synchronization}), the two notions \emph{are} equivalent: the MSF equals the largest transverse Floquet exponent of the attractor (Theorem 2.83 in Ref.~\cite{chicone2006ordinary}); thus, a strictly negative MSF implies transverse asymptotic stability (and vice versa) in that case, both with the original formulation of MSF \cite{pecora1998master} and its extension \cite{nishikawa2006synchronization,nishikawa2006maximum}.

For the specific periodic attractor studied in Ref.~\cite{muolo2021synchronization}, which we denote by $\Gamma := \{ x_1 = \cdots = x_n = x \in \Gamma^* \}$, where $\Gamma^* \subset \mathbb{R}^2$ is the periodic trajectory of a single oscillator, the strictly negative MSF values verified in Fig.~5b imply the transverse asymptotic stability of $\Gamma$.
It then follows that the basin of attraction of $\Gamma$ in each section transverse to the synchronization manifold contains an open neighborhood of $\Gamma$, and thus there exists a $\delta > 0$ such that the system converges back to $\Gamma$ under any transverse perturbation with magnitude $< \delta$, regardless of the location on $\Gamma$ where the perturbation is applied.
In general, the \emph{instantaneous} growth/decay rates of such perturbations can vary with the position along $\Gamma$, but it is the \emph{average} (exponential) growth/decay rates over the entire $\Gamma$ that is characterized by the Floquet exponents (and the MSF).
Since the attraction basin of $\Gamma$ contain an open neighborhood of it, the basin boundary cannot touch $\Gamma$, meaning that instability cannot arise if the perturbation is sufficiently small, even in the nonlinear regime.
However, instabilities \emph{do} arise when finite-size perturbations are large enough to cross the basin boundary.
Thus, the transverse asymptotic stability of $\Gamma$ implied by the strictly negative MSF is by no means in contradiction with other numerics in Ref.~\cite{muolo2021synchronization} showing finite-size perturbations that induce desynchronization instabilities.
None of the results in the paper can therefore be used to claim any failure of the MSF technique.

The caption of Fig.~4  discusses the ``mechanism that drives the instability in the non-normal linearized regime'' \cite{muolo2021synchronization}. However, in the linear regime, any `instability' must be transient because the synchronization is transversely asymptotically stable according to the MSF analysis.
It is true that in chaotic systems exhibiting bubbling \cite{ashwin1994bubbling,heagy1995desynchronization,venkataramani1996transitions} instabilities can be induced by infinitesimal perturbations to specific atypical synchronization trajectories even when the MSF for typical trajectories is negative.
But, as already noted, a similar effect is not possible for the periodic synchronization trajectory considered in Ref.~\cite{muolo2021synchronization}.
Moreover, when bubbling does occur, it is generally due to trajectories embbeded in the chaotic attractor (typically periodic trajectories with low periods) that are transversely unstable and hence have a positive MSF (despite the negative MSF of typical trajectories on the attractor).
Therefore, as established in Ref.~\cite{pecora1998master}, the MSF for synchronization stability analysis is applicable even to bubbling scenarios provided that the relevant trajectories are accounted for.
The same holds for the extension introduced in Refs.~\cite{nishikawa2006synchronization,nishikawa2006maximum}, and thus the results established therein on optimal networks are applicable to not only periodic but also chaotic synchronization dynamics.

\subsection*{Additional issues}

Ref.~\cite{muolo2021synchronization} also includes other invalid statements, of which the following are  relevant in connection with the points clarified above:

\begin{itemize}

\item
It states that ``a high non-normality translates to a high spectral degeneracy.'' This is not correct because there are many networks whose Laplacian matrix is highly non-normal but has no spectral degeneracy.

\item 
It also states that the directed Erd\H{o}s-R\'{e}nyi networks 
introduced in Appendix A  ``will be automatically non-normal,'' but this is not correct, as they are not guaranteed to be non-normal in general.

\item 
Likewise, the claim that for ``a given threshold of the control parameter $p$, the networks become degenerate'' (signified by a sharp transition from a solid to dashed line in Fig.~A1a) does not hold. Even though the likelihood of degeneracy increases with the connection probability $p$, the fraction of degenerate networks must be nonzero and strictly smaller than one for any~$p$.

\end{itemize}

\bibliography{refs}% Produces the bibliography via BibTeX.

\end{document}